\documentclass[reprint,superscriptaddress,aps,twocolumn,showpacs]{revtex4-2}
\usepackage{amssymb}
\usepackage{amsmath}
\usepackage[dvips]{graphicx}
\usepackage{hyperref}
\usepackage{braket,mleftright}
\usepackage{xcolor}
\usepackage{bm}

\allowdisplaybreaks

\begin{document}

\twocolumngrid

\title{Investigation of $\Delta(1232)$ resonance substructure in $p\gamma^*\to \Delta(1232)$ process through helicity amplitudes}
	
\author{A.~Kaewsnod}
\email[]{a.kaewsnod@gmail.com}
\author{K.~Xu}
\author{T.~Sangkhakrit}
\author{Z.~Zhao}
\author{W.~Sreethawong}
\author{A.~Limphirat}
\email[]{ayut@g.sut.ac.th}
\author{K.~Khosonthongkee}
\author{Y.~Yan}
\email[]{yupeng@g.sut.ac.th}
\affiliation{School of Physics and Center of Excellence in High Energy Physics and Astrophysics, Suranaree University of Technology, Nakhon Ratchasima 30000, Thailand}
	
\date{\today}
	
\begin{abstract}
This work investigates the substructure of the $\Delta(1232)$ resonance in the $p\gamma^*\to \Delta(1232)$ process through helicity transition amplitudes within the quark model framework.
We consider the involved baryons composed of three quarks, and both the quark core and meson cloud contribute to the transition amplitudes.
The comparison of theoretical results with experimental data reveals that, rather than the  $L=0$ component of the $\Delta(1232)$ resonance, it is the $L=2$ component that significantly affects its $S_{1/2}$ amplitude.
These findings indicate that the $\Delta(1232)$ resonance likely contains a substantial $L=2$ component, challenging the conventional view of the $\Delta(1232)$ resonance as an $L=0$ baryon.

\end{abstract}
	
\maketitle
	
\section{\label{sec:Int} Introduction}

The investigation of the mass spectra and internal structure of baryon resonances, particularly those described by three-quark models, has been an active area of research for several decades.
Among these resonances, states such as $N(1440)$, $N(1520)$, and $N(1535)$ are of particular interest due to persistent discrepancies between theoretical predictions and experimental results \cite{Tiator2011,Tiator2009,Aznauryan2007,Aznauryan2008,Capstick2007,Ramalho2010,Golli2009,Burkert2019,Li2006,Obukhovsky2011,Cano1998,Stajner2017}.
These challenges highlight limitations in traditional quark models and underscore the need for incorporating additional dynamical components, such as meson-baryon and multiquark configurations \cite{Sarantsev2008,Capstick2000}.

One promising approach to resolving these discrepancies involves considering pentaquark and meson cloud contributions in baryonic models.
Recent studies suggest that these non-conventional components are critical for understanding the resonance spectrum more comprehensively \cite{PhysRevD.101.076025}.
In particular, investigations into the helicity transition amplitudes have provided new insights into the structure of resonances like $N(1440)$, $N(1520)$, and $N(1535)$, revealing potential admixtures of higher orbital momentum and exotic quark configurations.

Meson cloud contributions have been shown to significantly affect both the transition form factors and helicity amplitudes of several nucleon excitations.
For example, in \cite{PhysRevD.108.074019}, the kaon cloud was found to enhance the estimates of radiative decays of strange baryons like $\Sigma(1385)$ and $\Xi(1530)$.
This study emphasizes that while the pion cloud dominates processes like the $\gamma^*N \rightarrow \Delta(1232)$ transition, the kaon cloud plays an essential role in strange baryons.
These contributions extend to the Dalitz decays, demonstrating the profound impact of mesonic degrees of freedom on the form factors of baryon resonances.

Further insight into the meson cloud’s role in the $\gamma^*N \to \Delta(1232)$ transition is provided by \cite{Ramalho2019}, which presents a global parametrization of the neutron electric form factor and the quadrupole form factors, $G_E$ and $G_C$.
The study highlights the dominance of the pion cloud in these transitions, particularly at low $Q^2$, and emphasizes that the quadrupole form factors exhibit large radii due to the extended structure of the pion cloud.
These results suggest that meson cloud effects, especially from the pion, are essential for accurately describing the $\Delta(1232)$ resonance.

In \cite{Ramalho2018}, the meson cloud's relevance at moderate $Q^2$ values is further explored, particularly for resonances such as $N(1535)$ and $N(1520)$.
The study shows that while valence quark effects dominate at higher $Q^2$, the meson cloud significantly influences the form factors and helicity amplitudes at lower $Q^2$, underscoring its importance in the non-perturbative regime of QCD.

The study of $D$-wave state admixtures in the wave functions of nucleons and $\Delta(1232)$ has also provided valuable insight.
It has been demonstrated that while $D$-wave state components have minimal effects on the elastic form factors of the proton and the magnetic form factor of the neutron, they significantly influence the electric form factor of the neutron \cite{JULIADIAZ2005441}.
Moreover, in the $\Delta(1232)-N$ transition, the $D$-wave state admixtures strongly affect the electric form factor, particularly in the ratio $R_{EM}$, which is highly sensitive to these components.
These observations suggest that $D$-wave state admixtures are essential for understanding the full structure of the $\Delta(1232)$ resonance.

Further, the inclusion of $D$-wave state components in the $\Delta(1232)$ wave function has been shown to be necessary for describing the electric quadrupole and magnetic octupole form factors of the $\Delta$ resonance.
These contributions are key to explaining the $\gamma^* N \to \Delta$ quadrupole transition and provide a robust framework for analyzing lattice QCD results \cite{PhysRevD.81.113011}.
At high $Q^2$ values, the valence quark contributions dominate, while at low $Q^2$, pion cloud effects become more relevant.

In this context, the $\Delta(1232)$ resonance stands out as a key subject of interest due to its complex internal structure.
While the $\Delta(1232)$ is often described as a predominantly $S$-wave three-quark ($q^3$) state, experimental data reveal non-zero longitudinal helicity transition amplitudes ($S_{1/2}$) for the $\gamma^* N \rightarrow \Delta(1232)$ process.
This observation implies that the $\Delta(1232)$ may not be a pure $S$-wave state and could involve significant higher orbital ($D$-wave state) components, as well as contributions from meson cloud effects and possible multiquark admixtures \cite{Kaewsnod2022a,Kaewsnod2022b}.

The present study focuses on the substructure of the $\Delta(1232)$ resonance in the $p\gamma^* \rightarrow \Delta(1232)$ process, using helicity amplitudes as a key probe.
We apply a quark model framework that incorporates both the quark core and meson cloud contributions, considering baryons as composite three-quark systems.
By analyzing the helicity amplitudes of low-lying resonances such as $\Delta(1232)$, $N(1440)$, $N(1520)$, and $N(1535)$, this work aims to provide deeper insights into the underlying dynamics, paving the way for future studies on higher resonances like $N(1875)$ and $N(1895)$.

The structure of this paper is as follows.
In Sec.~\ref{sec:WF}, we present the wave functions of the proton and $\Delta(1232)$ utilized in this study.
In Sec.~\ref{sec:Helicity amplitudes}, the helicity amplitudes $A_{1/2}$, $A_{3/2}$, and $S_{1/2}$ for the $p\gamma^*\to\Delta(1232)$ transition are introduced, along with the corresponding quark-core and meson-cloud diagrams, depicted in Figs.~\ref{fig:quarkcore} and \ref{fig:MC}, respectively.
The results of the helicity amplitudes, incorporating contributions from both diagrams in Sec.~\ref{sec:Helicity amplitudes}, are presented in Sec.~\ref{sec:Result}.
Finally, conclusions are summarized in Sec.~\ref{sec:Summary}.

\section{\label{sec:WF}Proton and $\Delta(1232)$ in a Three-Quark Model}

The algebraic structure of the proton and $\Delta(1232)$ baryons is governed by the product of standard spin, flavor, and color symmetries: $SU_f(2) \otimes SU_s(2) \otimes SU_c(3)$.
We propose that the spatial wave function of the $\Delta(1232)$ consists of a fully symmetric component in the $L = 0$ state, along with two additional components in the $L = 2$ state—one symmetric (Spin $S=\frac{3}{2}$) and one of mixed symmetry ($S=\frac{1}{2}$).
The wave functions of the proton and $\Delta(1232)$ within the three-quark model can be expressed as:
\begin{alignat}{1}
\Psi_p = &\,\frac{1}{\sqrt 2}\psi^{S}_{p,0}\left( \chi^\lambda\phi^\lambda+\chi^\rho\phi^\rho\right)\psi^c,  \nonumber\\
\Psi_\Delta = &\,\bigg[  \, B\,\psi^{S}_{\Delta,0}\chi^S +C\,\psi^{S}_{\Delta,2} \otimes \chi^S  \nonumber \\
& + D\,\frac{1}{\sqrt 2} \Big( \psi^{\lambda}_{\Delta,2} \otimes \chi^\lambda + \psi^{\rho}_{\Delta,2} \otimes \chi^\rho \Big) \bigg] \phi^S \psi^c.
\label{eq:p-Delta WF}
\end{alignat}
where $\psi^c$ is the color wave function, and $\chi^\alpha$ and $\phi^\alpha$ are spin and flavor wave functions, respectively, of symmetry type $\alpha$.
The spatial wave functions \(\psi^{\alpha}_{X,L}\) describe the \(X\) baryon with orbital angular momentum \(L\) and symmetry \(\alpha\), where \(\alpha = S\) (symmetric), \(\lambda\) (lambda-type), or \(\rho\) (rho-type).

The spatial wave functions for the proton and $\Delta(1232)$ baryons are represented in a harmonic oscillator basis using Jacobi coordinates, as detailed in Refs.~\cite{PhysRevC.100.065207,PhysRevD.101.076025}. These wave functions are expressed as:
\begin{align}
\psi^S_{p,0} = &\ \sum_{N=0,2,4,\dots} a_N \, \Psi_{N00}^S,\nonumber\\
\psi^{S}_{\Delta,0} = &\ \sum_{N=0,2,4,\dots} b_N \, \Psi_{N00}^{S}\nonumber\\
\psi^{S}_{\Delta,2} = &\ \sum_{N=2,4,6,\dots} c_N \, \Psi_{N2M}^{S}\nonumber\\
\psi^{\lambda/\rho}_{\Delta,2} = &\ \sum_{N=2,4,6,\dots} d_N \, \Psi_{N2M}^{\lambda/\rho}.
\end{align}
These are subject to the following normalization conditions:
\begin{align}
B^2 + C^2 + D^2 = 1,\nonumber\\
\sum_{n=0,2,4,\dots} \left(a_N\right)^2 = \sum_{N=0,2,4,\dots} \left(b_N\right)^2 = 1,\nonumber\\
\sum_{N=2,4,6,\dots} \left(c_N\right)^2 = \sum_{N=2,4,6,\dots} \left(d_N\right)^2 = 1.
\end{align}
Here, $\Psi_{NML}^{\alpha}$ represents the harmonic oscillator basis in Jacobi coordinates, with $N$ as the principal quantum number, $L$ as the angular momentum, and $M$ as the angular momentum projection. The symmetry type of the baryon is denoted by $\alpha$.
The constant $a_N$ in the spatial wave function of the proton is obtained by fitting to experimental data from its electric form factor (see Ref.~\cite{Kaewsnod2022a} for more details).
For the \(\Delta(1232)\) resonance, the constants \( B \), \( C \), \( D \), \( b_N \), \( c_N \), and \( d_N \) in its spatial wave function are obtained through fitting to experimental data on its helicity amplitudes.
The results of this fitting are presented in Section \ref{sec:Result}.
	
\section{\label{sec:Helicity amplitudes} Helicity amplitudes with $\Delta(1232)$ as three quark states}
We assumed that both the proton $p$ and the resonance $\Delta$ are composed of three quarks.
The transverse helicity amplitudes $A_{1/2}$ and $A_{3/2}$, and the longitudinal helicity amplitude $S_{1/2}$ are usually defined in the resonance rest frame as

\begin{figure}[t!]
	\centering
	\setlength{\abovecaptionskip}{-5pt} 
	\includegraphics[width=0.4\textwidth]{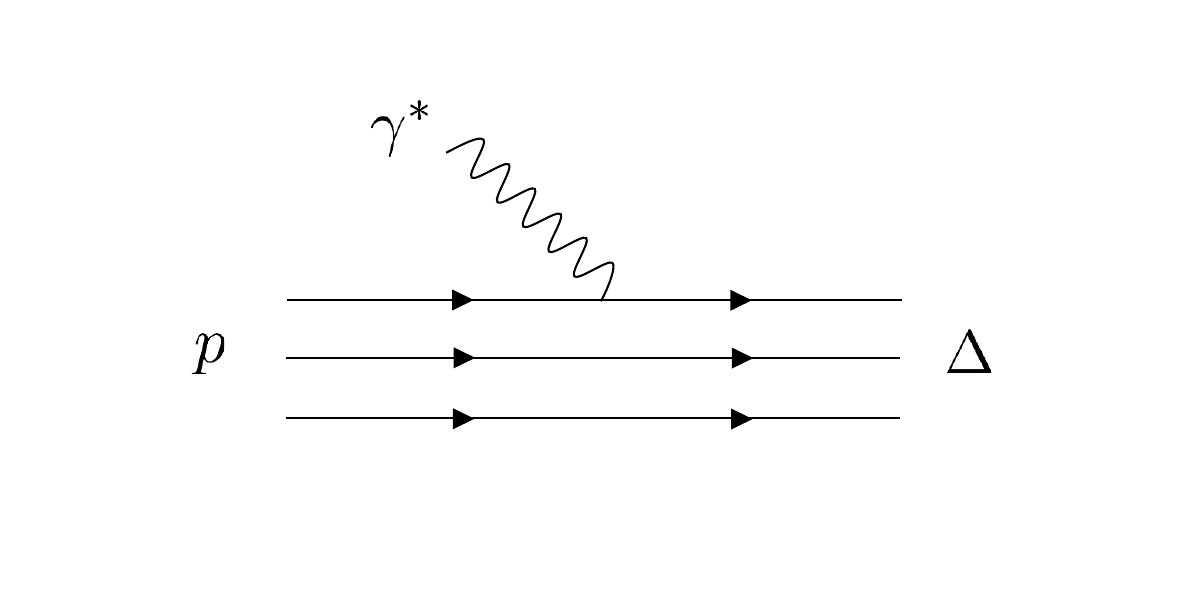}
	\caption{Diagram of the $p\gamma^* \to \Delta(1232)$ transition, where the photon directly interacts with the quark core of the proton, leading to the production of a $ \Delta(1232)$ resonance through the photoproduction process.}
	\label{fig:quarkcore}
\end{figure}

\begin{figure}[t!]
	\centering
	\setlength{\abovecaptionskip}{-5pt} 
	\includegraphics[width=0.4\textwidth]{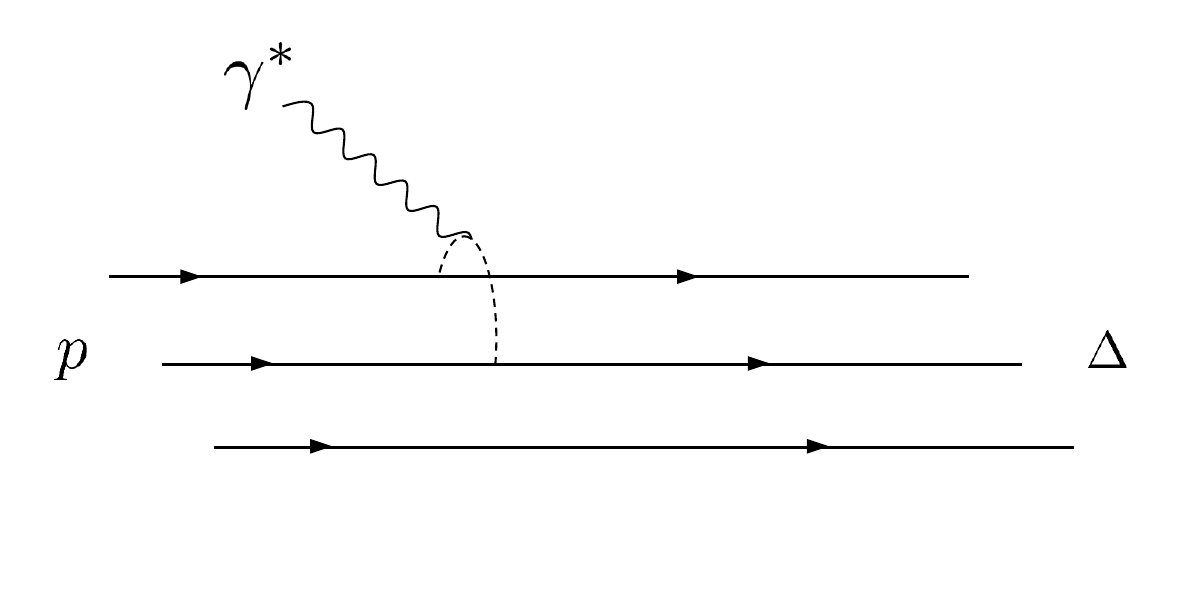}
	\caption[]{Similar to the diagram in FIG.~\ref{fig:quarkcore}, but in this case, the photon first interacts with the meson cloud, which then interacts with the quark inside the proton, leading to the photoproduction of the $\Delta(1232)$ resonance.}
	\label{fig:MC}
\end{figure}

\onecolumngrid
\begin{widetext}

\begin{alignat}{5}
&A_{1/2}&&=&&\
\dfrac{1}{\sqrt{2K}}\Braket{\Delta,S'_z=1/2|q_1'q_2'q_3'}\left( T_{q^3}^++T_{MC}^+ \right) \Braket{q_1q_2q_3|p,S_z=-1/2},\nonumber\\
&A_{3/2}&&=&&\ \dfrac{1}{\sqrt{2K}}\Braket{\Delta,S'_z=3/2|q_1'q_2'q_3'}\left( T_{q^3}^++T_{MC}^+ \right)\Braket{q_1q_2q_3|p,S_z=1/2},\nonumber\\
&S_{1/2}&&=&&\ \dfrac{1}{\sqrt{2K}}\Braket{\Delta,S'_z=1/2|q_1'q_2'q_3'}\left( T_{q^3}^0+T_{MC}^0 \right)\Braket{q_1q_2q_3|p,S_z=1/2}.
\end{alignat}


\end{widetext}
\twocolumngrid
The real photon momentum is represented as $K=\frac{M_{\Delta}^2-M_p^2}{2M_{\Delta}}$ where $ M_{\Delta}$ and $M_p$ refer to the rest masses of the $\Delta(1232)$ resonance and the proton, respectively.
$T^\lambda_{q^3}$ represents the transition corresponding to the interaction between the photon and quark $i$ in the electromagnetic process $\gamma q_i \to q_i'$, as illustrated by the quark line diagram in Fig.~\ref{fig:quarkcore}.
$T^\lambda_{MC}$ denotes the transition where the photon interacts with the meson cloud, which subsequently couples to quark $i$, as depicted in the quark line diagram in Fig.~\ref{fig:MC}.
The superscript $\lambda$ refers to the helicity, where $\lambda = 0$ represents the longitudinal helicity amplitude, and $\lambda = +1$ corresponds to the transverse helicity amplitude.
$\Braket{p,S_z|q_1q_2q_3}$ and $\Braket{\Delta,S'_z|q_1'q_2'q_3'}$ represent the three-quark wave functions of the initial proton state and the final $\Delta(1232)$ resonance state, respectively, with spin projections \(S_z\) for the proton and \(S'_z\) for the $\Delta$.
The calculation considers all possible intermediate states of the three quarks in the three-quark picture, which requires summing over spin, flavor, and color, and integrating over all quark momenta.
For more details, we refer the reader to our previous works about the study of $N(1440)$, $N(1520)$, and $N(1535)$ \cite{Kaewsnod2022a,Kaewsnod2022b}.

The electromagnetic transition of the quark core contribution is approximated using the impulse approximation.
In this approximation, the interaction of the photon with one quark is described by the coupling of the quark current to the electromagnetic field in the resonance rest frame.
For instance, when the photon interacts with the third quark, we have
\begin{alignat}{2}
T^\lambda_{q^3} =\ Q_3\bar u_{s'}(p')\gamma^\mu u_s(p)\epsilon^\lambda_\mu(k)\Braket{q_1q_2|q_1'q_2'}.
\end{alignat}
Here, $Q_3$ represents the electric charge of the third quark, $u_s(p)$ is the Dirac spinor with spin $s$ and momentum $p$, the polarization vector of the photon ($\epsilon^\lambda_\mu$) is defined as $\epsilon^+_\mu = (0, -1, -i/\sqrt{2}, 0)$ and $\epsilon^0_\mu = (\frac{K}{Q}, 0, 0, \frac{\omega}{Q})$ where $\omega$ is a photon energy.
The transition amplitude for the longitudinal helicity component, $S_{1/2}$, is evaluated following the formalism described in Refs.~\cite{Dreschsel_1992, DRECHSEL1999145, Drechsel2007}.
The $p$ wave function $\Braket{q_1q_2q_3|p,S_z}$ and $\Delta(1232)$ wave function $\Braket{q_1'q_2'q_3'|\Delta,S'_z}$ in the three-quark picture have been shown in Eq.~\ref{eq:p-Delta WF}.
In this work, we adopt a current quark mass for the up ($u$) and down ($d$) quarks, setting $m = 5$ MeV.

The transition involving a photon interacting with a quark via a meson cloud can be approximated by considering only the leading-order contribution, as illustrated by the quark line diagram in Fig.~\ref{fig:MC}.
Higher-order corrections are assumed to be suppressed and are neglected under this approximation.
Within the framework of the chiral quark model, the dynamics of the interaction are governed by the following effective Lagrangians \cite{Liu_2014}
\begin{align}
\mathcal{L}_{qq\pi}=&\frac{1}{2F}  \partial_\mu \pi_i \bar\psi \gamma^\mu \gamma^5 \tau^i \psi
\nonumber\\
\mathcal{L}_{\pi\pi\gamma}= &-e \,\epsilon_{3jk}\pi_j \partial^\nu  \pi_k A_\nu,
\end{align}
where the first term $\mathcal{L}_{qq\pi}$ describes the interaction between quarks and a pion, with $F=88$ MeV representing the pion decay constant, $\tau^i$ the Pauli matrices acting in isospin space, and $\pi_i$ the pion fields.
The second term $\mathcal{L}_{\pi\pi\gamma}$ represents the interaction between a photon and pion fields, where $A_\nu$ is the photon field, and $\epsilon_{3jk}$ is the Levi-Civita symbol that enforces the symmetry properties of the pion-photon interaction.

\begin{figure}[t]
	\centering
	\includegraphics[width=0.4\textwidth]{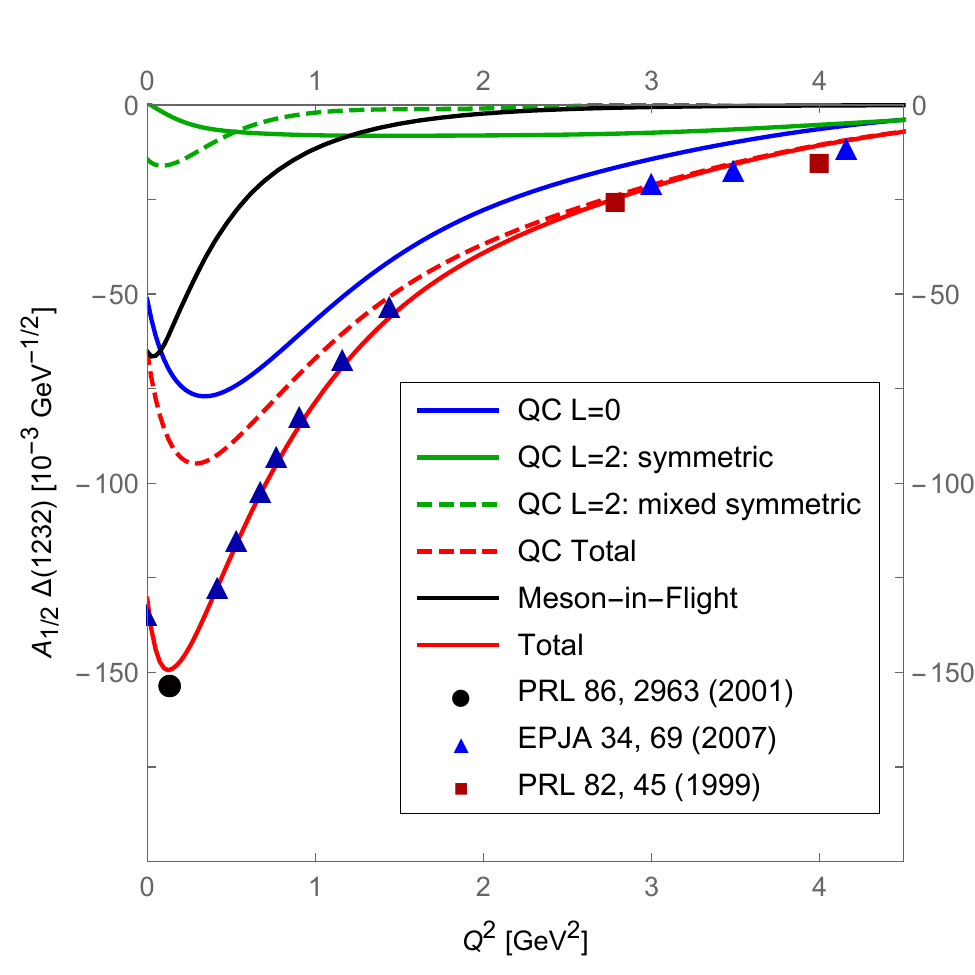}
	\caption[$A_{1/2}$ of $\Delta(1232)$]{Transverse helicity amplitude $A_{1/2}$ for the $p\gamma^* \to \Delta(1232)$ transition compared to experimental data.
		The blue and green curves represent quark core (QC) contributions: blue (symmetric $L=0$), green solid (symmetric $L=2$), and green dashed (mixed-symmetric $L=2$) components of the $\Delta(1232)$ wave function.
		The black curve shows the pion cloud contribution.
		The red dashed curve combines the $L=0$ and $L=2$ components, while the red solid curve presents the total result, including both quark core and pion cloud contributions.
	The experimental data are taken from \cite{PhysRevLett.86.2963, Drechsel2007, PhysRevLett.82.45}.}
	\label{fig:A12}
\end{figure}

\section{\label{sec:Result} Results}
\begin{figure}[t]
	\centering
	\includegraphics[width=0.4\textwidth]{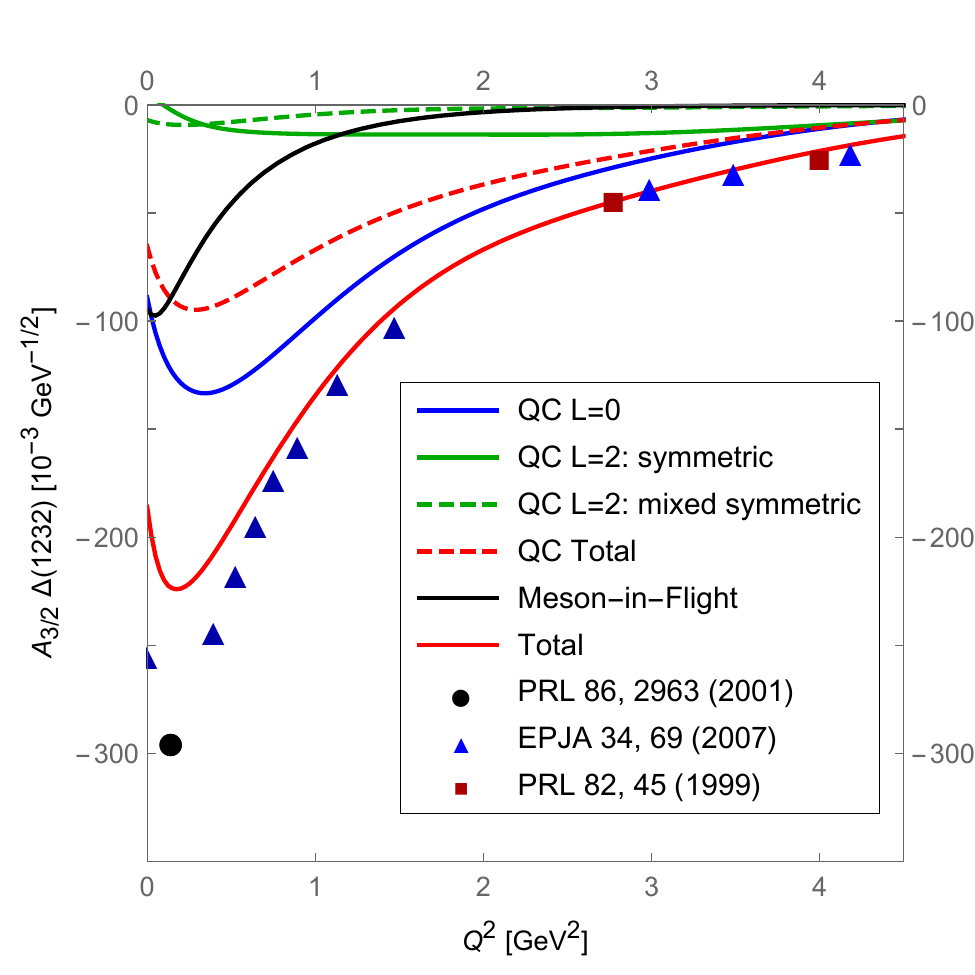}
	\caption[$A_{3/2}$ of $\Delta(1232)$]{Transverse helicity amplitude $A_{3/2}$ of the $p\gamma^* \to \Delta(1232)$ transition compared to experimental data.
	The notations for the curves are the same as those described in Fig.~\ref{fig:A12}.}
	\label{fig:A32}
\end{figure}

\begin{figure}[t]
	\centering
	\includegraphics[width=0.39\textwidth]{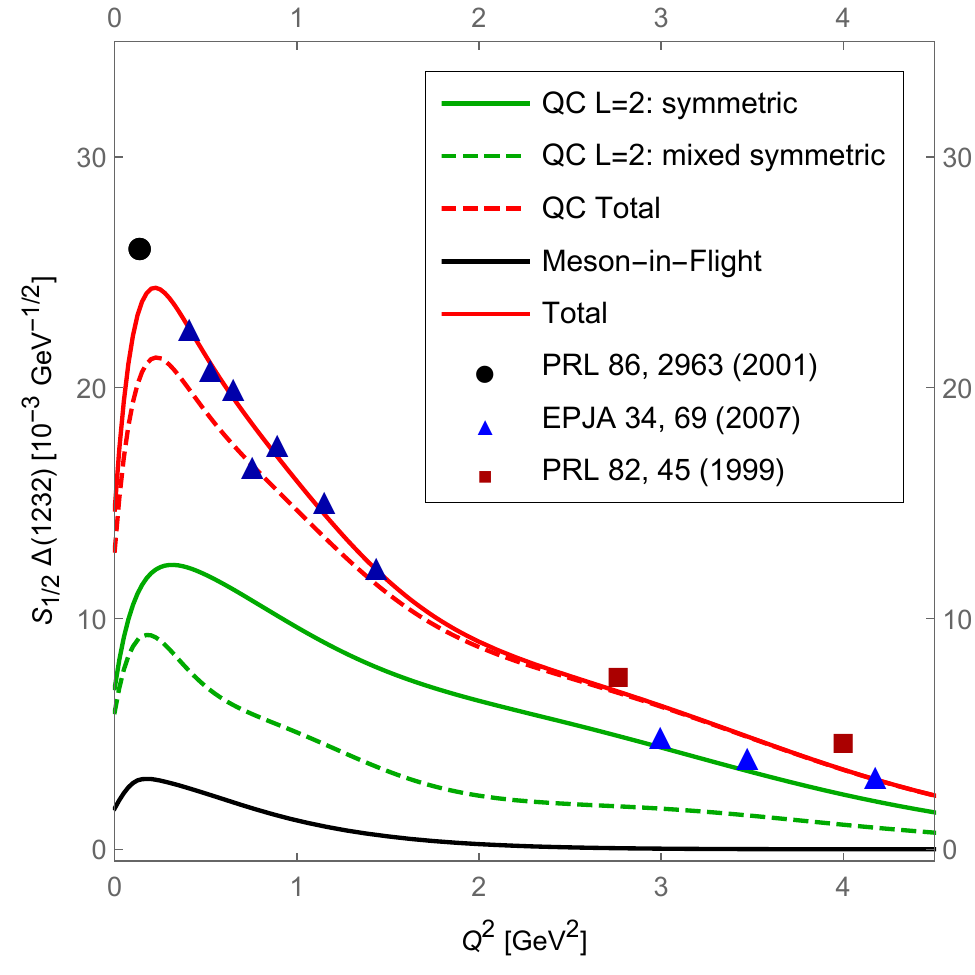}
	\caption[$S_{1/2}$ of $\Delta(1232)$]{Longitudinal helicity amplitude $S_{1/2}$ of the $p\gamma^* \to \Delta(1232)$ transition compared to experimental data.
	The notations for the curves are the same as those described in Fig.~\ref{fig:A12}.}
	\label{fig:S12}
\end{figure}

The Figs.~\ref{fig:A12}, \ref{fig:A32}, and \ref{fig:S12} present the helicity amplitudes of the $\Delta(1232)$ resonance from the quark-core diagram and meson-cloud contribution compared to the experimental data (shown as colored dots).
The wave function of the $\Delta(1232)$ resonance is considered with both orbital angular momenta $L=0$ and $L=2$ , while the proton's wave function is extracted from its electric form factor with $L=0$.
In the meson cloud diagram shown in Fig.~\ref{fig:MC}, a monopole form factor is introduced for the pion–photon interaction. The associated cutoff parameter, $\Lambda_\pi = 0.732$ GeV, is selected to be consistent with the empirical pion charge form factor. It is worth noting that the meson cloud contribution is most significant in the low-$Q^2$ region of the resonance.

The comparison between theoretical predictions and experimental data indicates that the $\Delta(1232)$ resonance consists of 53\% ($B = 0.725$) in the $L=0$ state, 23\% ($C = 0.485$) in the $L=2$ symmetric state, and 24\% ($D = 0.487$) in the $L=2$ mixed-symmetric state.
Particularly at higher momentum transfers, the result reveals that the $\Delta(1232)$ may comprise a considerable $L=2$ component.
Specifically, in the $S_{1/2}$, the absence of contribution from $\Delta(1232)$ with $L=0$ is conspicuous; only the $L=2$ component appears to influence $S_{1/2}$.
This observation challenges the conventional notion that the $\Delta(1232)$ baryon primarily possesses an $L=0$ configuration, presenting a compelling avenue for further investigation and refinement of our understanding.


\section{\label{sec:Summary}Summary and Discussion}

In this paper, we have investigated the substructure of the $\Delta(1232)$ resonance in the $p\gamma^*\to \Delta(1232)$ process using helicity amplitudes.
Our approach utilizes the three-quark model, incorporating both quark core and meson cloud contributions, with the proton and $\Delta(1232)$ baryon wave functions constructed based on the symmetries of spin, flavor, and color ($SU_f(2) \otimes SU_s(2) \otimes SU_c(3)$).
The spatial wave functions are represented within a harmonic oscillator basis in Jacobi coordinates, with the $\Delta(1232)$ including both $L=0$ and $L=2$ components, while the proton remains in an $L=0$ state.
The coefficients for these wave function components were fitted to experimental data.

The helicity amplitudes $A_{1/2}$, $A_{3/2}$, and $S_{1/2}$, defined in the resonance rest frame, were computed by considering the electromagnetic transitions between quark states.
These transitions involve both direct photon-quark interactions, represented in the quark core diagram, and interactions via the meson cloud.
The meson cloud contributions were described using the chiral quark model, where we adopted a monopole form factor for the pion-photon interaction to handle loop singularities and applied Pauli–Villars regularization for the pion-quark interaction.
The results show that the meson cloud has a significant influence, particularly in the low-$Q^2$ region.

From our results, the $\Delta(1232)$ resonance was found to consist predominantly of the $L=0$ component, which makes up 53\% of its structure, with smaller contributions from the symmetric ($L=2$, 23\%) and mixed-symmetric ($L=2$, 24\%) components.
These results challenge the traditional understanding that the $\Delta(1232)$ baryon is purely an $L=0$ state.
Particularly in the longitudinal helicity amplitude $S_{1/2}$, the $L=0$ component was shown to have no contribution, with only the $L=2$ component playing a role.
This suggests that higher angular momentum states may be more relevant to the $\Delta(1232)$ resonance than previously thought.

The agreement between our theoretical predictions and experimental data, as demonstrated in Figs.~\ref{fig:A12}, \ref{fig:A32}, and \ref{fig:S12}, confirms the importance of including both quark core and meson cloud contributions to fully describe the helicity amplitudes.
The presence of a significant $L=2$ component in the $\Delta(1232)$ resonance structure opens up intriguing possibilities for future studies, particularly in refining models that go beyond the simple quark core picture.
Our findings also support the necessity of considering meson cloud effects in baryon resonance transitions, providing valuable insights into the dynamics of the $p\gamma^*\to \Delta(1232)$ process.

Further exploration of higher-order corrections to the meson cloud and quark core interactions, as well as a more detailed investigation of the role of $L=2$ states in other baryon resonances, would be important next steps.
This work lays a foundation for these future studies and contributes to the ongoing effort to understand the internal structure of baryons in terms of quarks and mesons.

\section*{ACKNOWLEDGMENTS}
This research has received funding support from (i) Suranaree University of Technology (SUT) and (ii) the NSRF via the Program Management Unit for Human Resources \& Institutional Development, Research and Innovation (PMU-B) [grant number B13F660067]
	
%
%
	
\bibliographystyle{unsrtnat}
\bibliography{bibtex}
	
\end{document}